\documentclass[slac_one]{revtex4}
\usepackage{graphicx}
\usepackage{fancyhdr}
\def\stacksymbols #1#2#3#4{\def\theguybelow{#2}
    \def\vp{\lower#3pt}
    \def\sp{\baselineskip0pt\lineskip#4pt}
    \mathrel{\mathpalette\intermediary#1}}

\def\intermediary#1#2{\vp\vbox{\sp
     \everycr={}\tabskip0pt
     \halign{$\mathsurround0pt#1\hfil##\hfil$\crcr#2\crcr
              \theguybelow\crcr}}}

\def\lapproxeq{\stacksymbols{<}{\sim}{2.5}{.2}}

\def\beq{\begin{equation}}
\def\eeq#1{\label{#1}\end{equation}}
\def\eeqn{\end{equation}}
\def\beqa{\begin{eqnarray}}
\def\eeqa#1{\label{#1}\end{eqnarray}}
\def\eeqan{\end{eqnarray}}

\def\leqn#1{(\ref{#1})}

\def\draftnote#1{{\bf [#1]}}

\begin{document}

\hfill$\vcenter{
\hbox{\bf MADPH-05-1251}
\hbox{\bf SLAC-PUB-11589}}
$

\title{{\small{2005 ALCPG \& ILC Workshops - Snowmass,
U.S.A.}}\\ 
\vspace{12pt}
Top Quark Properties in Little Higgs Models}

\author{C.~F.~Berger}
\affiliation{Stanford Linear Accelerator Center, Stanford
University, Stanford, CA 94309, USA}
\author{M.~Perelstein}
\affiliation{CIHEP, Cornell University, Ithaca, NY 14853, USA}
\author{F.~Petriello}
\affiliation{University of Wisconsin, Madison, WI 53706, USA}

\begin{abstract}
We study the shifts in the gauge couplings of the top quark induced in
the Littlest Higgs model with and without T parity. We find that the ILC
will be able to observe the shifts throughout the natural range of model
parameters.
\end{abstract}

\maketitle

\thispagestyle{fancy}

\section{INTRODUCTION}

Identifying the mechanism which breaks electroweak symmetry and generates
fermion masses is one of the main physics goals for both the LHC and the ILC.
Studies of the top quark have the potential to illuminate this issue;
since it is the heaviest of the Standard Model (SM) fermions, the top is
expected to
couple strongly to the symmetry-breaking sector. Consequently, the structure
of that sector can have significant, potentially observable effects on the
properties of the top. For example, it is well known that the vector and
axial $t\bar{t}Z$ form factors receive large corrections (of order 5-10\%)
in certain models of dynamical electroweak symmetry
breaking~\cite{snowmass01}. At future colliders such as the LHC and the ILC,
we will be able to pursue a program of
precision top physics, similar to the program studying the $Z$ at LEP and SLC.
In this manuscript, we study the corrections to
the top quark properties in ``Little Higgs'' models of electroweak symmetry
breaking~\cite{review}, and compare the expected deviations from the SM
 predictions with expected sensitivities of experiments at the LHC and
the ILC.

In the Little Higgs models, electroweak symmetry is driven by the
radiative effects from the top sector, including the SM-like top and its
heavy counterpart, a TeV-scale ``heavy top'' $T$. Probing this structure 
experimentally is quite difficult. While the LHC should be able to 
discover the $T$ quark, its potential for studying its couplings is 
limited~\cite{PPP,ATLAS}. Direct production of the $T$ will 
likely be beyond the kinematic reach of the ILC. However, we will show below
that the corrections to the gauge couplings of the SM top, induced by its
mixing with the $T$, will be observable at the ILC throughout the parameter 
range consistent with naturalness. Measuring these corrections will provide 
a unique window on the top sector of the Little Higgs.

\if
Little Higgs models contain a light Higgs boson which is a
{\it composite} of more fundamental 
degrees of freedom.
A generic composite Higgs model must become strongly coupled at an energy
scale around 1 TeV, leading to unacceptably large corrections to precision
electroweak observables. In contrast, Little Higgs models remain perturbative
until
a higher energy scale, around 10 TeV. The hierarchy between the Higgs mass
and the strong coupling scale is natural and stable with respect to radiative
corrections.  Because of the special symmetry structure of the theory,
the Higgs mass vanishes at tree level, as do one-loop
quadratically divergent diagrams.
The mass term is dominated by the
logarithmically divergent one-loop contribution from the top quark, which 
triggers electroweak symmetry breaking.
\fi

Many Little Higgs models have been proposed in the literature. We will
consider two examples in this study, the ``Littlest Higgs''
model~\cite{littlest}, and its variation incorporating T parity~\cite{LHT}.

\section{THE LITTLEST HIGGS}

As our first example, consider the $SU(5)/SO(5)$ Littlest Higgs (LH)
model~\cite{littlest}. Since the original model turned out to be severely
constrained by precision electroweak data~\cite{pew}, we focus on the version
with a
reduced gauge group, $SU(2)\times SU(2)\times U(1)$, which is significantly
less constrained. We follow the conventions and notation of Ref.~\cite{PPP}.
The new TeV-scale states are the gauge bosons $W_H^\pm, W_H^3$, a vector-like
weak-singlet quark $T$ (the ``heavy top''), and a weak-triplet scalar field
$\phi$. The model is parametrized by the symmetry breaking scale $f$ (assumed
to be of order 1 TeV), the $SU(2)$ mixing angle $\psi$, the $htT$ coupling
constant $\lambda_T$, and the triplet vacuum expectation value (vev)
$v^\prime$. It can be shown that $v^\prime\sim v^2/f\ll v$, where $v=246$
GeV is the SM Higgs vev. In this analysis, we will set $v^\prime=0$, since
the effects of a non-vanishing $v'$ on the observables considered here
are numerically small. Instead
of $f$, we will use the more physical
quantity, the heavy top mass $m_T$, in our plots; these are related by
$m_T/f = (\lambda_t^2+\lambda_T^2)/\lambda_T$, where $\lambda_t\approx 1$
is the SM top Yukawa. Naturalness arguments put an upper bound on this
parameter, $m_T\lapproxeq 2$ TeV.

\begin{figure*}[t]
\centering
\includegraphics[width=60mm,angle=90]{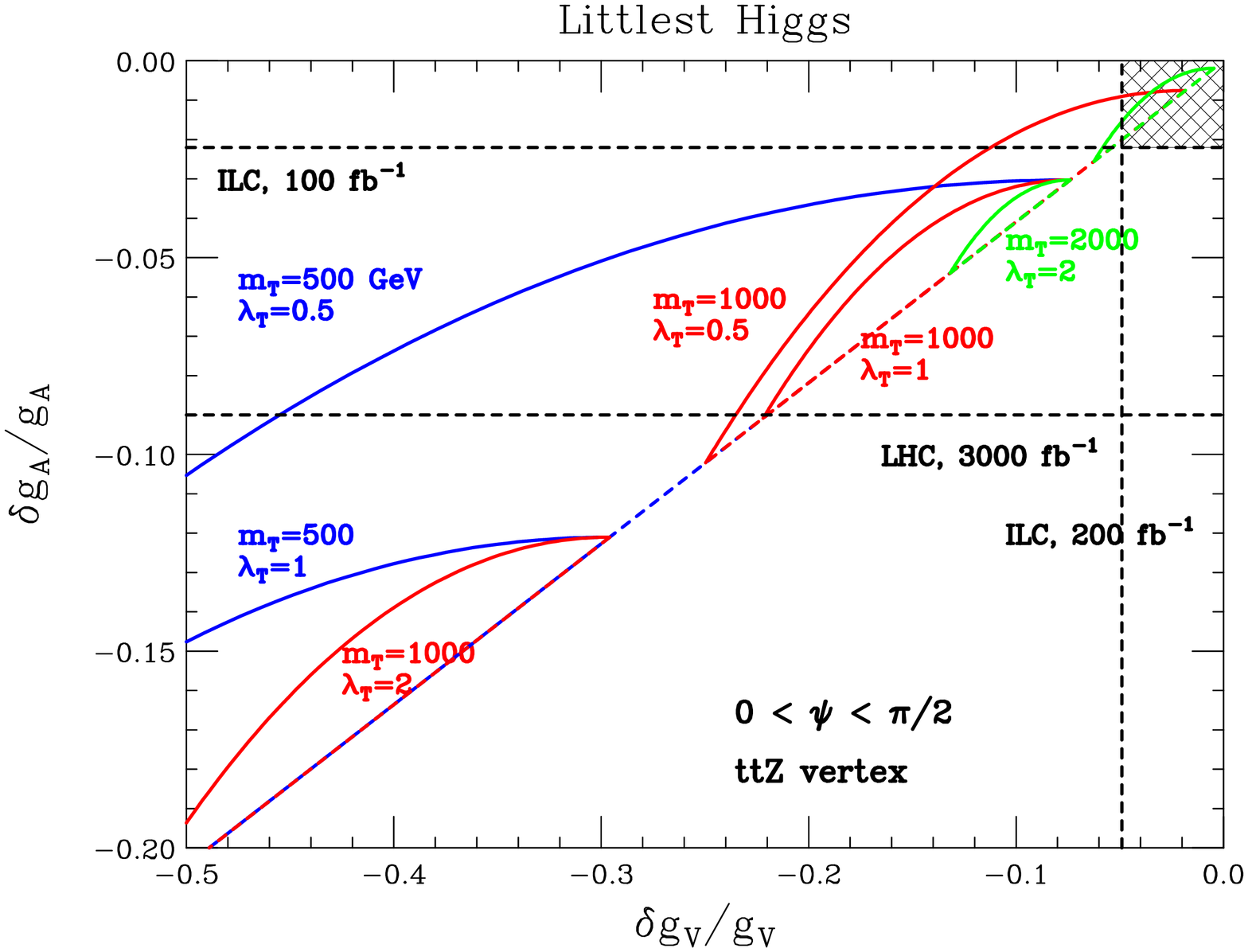}
\hskip1cm
\includegraphics[width=60mm,angle=90]{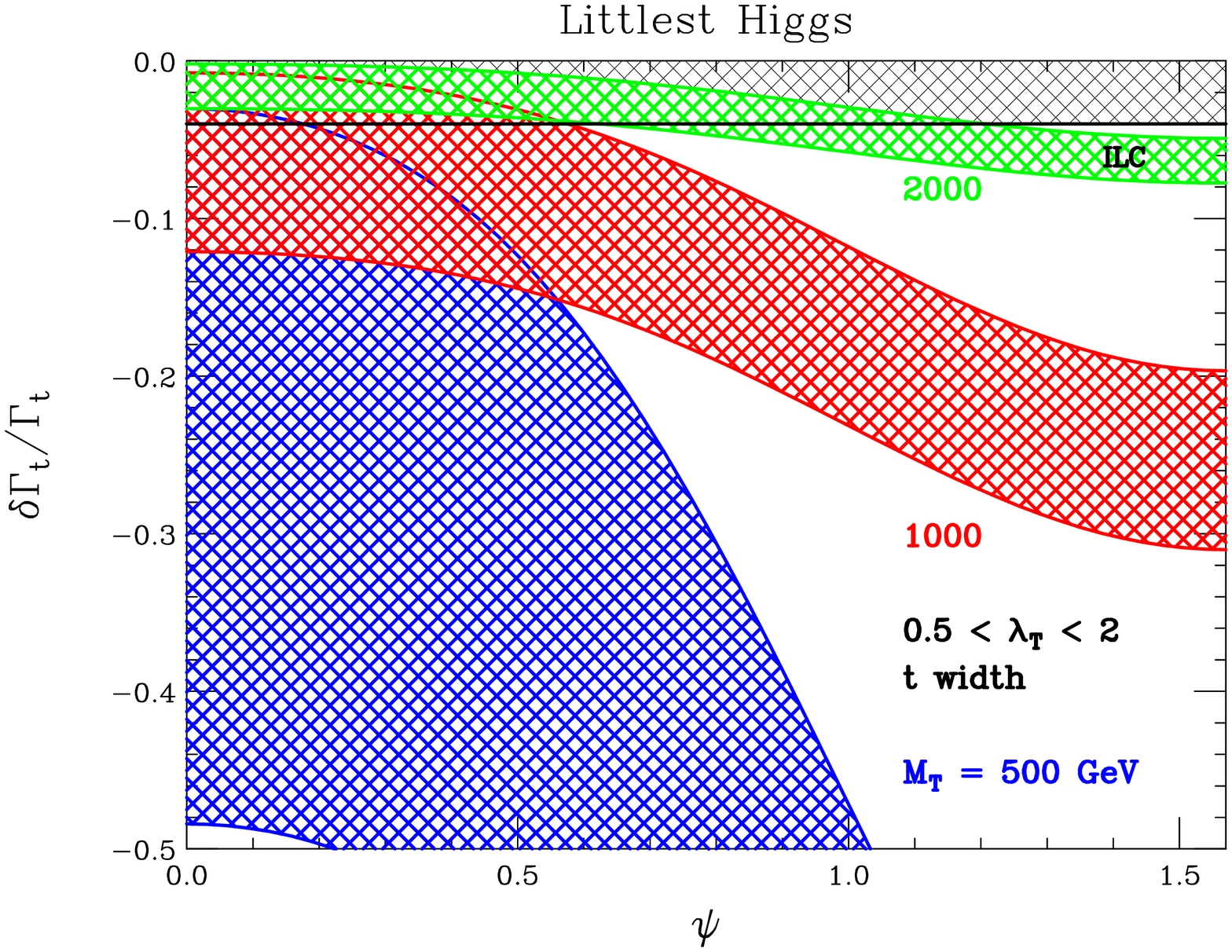}
\caption{The corrections to the $t\bar{t}Z$ axial and vector couplings (left
panel) and the top width $\Gamma_t$ (right panel) in the $SU(5)/SO(5)$
Littlest Higgs model. The regions in which the ILC would observe no 
deviation from the SM are shaded.}
\label{fig:LH}
\end{figure*}

Corrections to the gauge couplings of the top quark in the LH model arise
from two
sources: the mixing of the (left-handed) top with the heavy top $T$, and
the mixing of the SM gauge bosons $W^\pm, Z^0$ with their heavy counterparts,
$W^\pm_H$ and $W_H^3$. Using the superscripts ``t'' and ``g'' to denote the
contributions from these two sources, the corrections to the $t\bar{t}Z$
coupling can be written as
\beqa
\delta g^{Z{\rm t}}_R  = 0, ~~& &~~
\delta g^{Z{\rm g}}_R = \frac{v^2}{4 f^2}\frac{c_{\psi}^2 s_{\psi}^2}
{c_{W}^2-s_{W}^2}g_R^Z, \nonumber \\
\delta g^{Z{\rm t}}_L = \frac{\lambda_T^2 v^2 g_A^Z}{m_T^2}, ~~& &~~
\delta g^{Z{\rm g}}_L = \frac{v^2}{4 f^2}\left[2g_{A}^Z s_{\psi}^4+g_R^Z
\frac{c_{\psi}^2 s_{\psi}^2}{c_{W}^2-s_{W}^2}\right].
\eeqa{delta_ttZ}
Here, $g_{L,R}^Z$ are the SM left- and right-handed  $t\bar{t}Z$ couplings,
$g_V^Z=(g_R^Z+g_L^Z)/2$ and $g_A^Z=(g_R^Z-g_L^Z)/2$ are their vector and
axial combinations, $c_W,s_W$ are
respectively the cosine and sine of the weak mixing angle, and $s_\psi\equiv
\sin\psi$, $c_\psi\equiv\cos\psi$.
The predicted shifts in the $t\bar{t}Z$ axial and vector couplings
for $m_T=0.5, 1.0,$ and 2.0 TeV, and $\lambda_T=0.5, 1, 2$, are
plotted in Fig.~\ref{fig:LH} (left panel), along with the experimental
sensitivities expected at the LHC~\cite{LHC} and the ILC~\cite{snowmass01}.
The mixing angle $\psi$ is varied between $0$ and $\pi/2$.
Note that the shifts have a definite sign. While only a rather small part of
the parameter space is accessible at the LHC even with 3000 fb$^{-1}$
integrated luminosity, the ILC experiments will be able to easily observe the
shifts in most of the parameter space preferred by naturalness considerations.

The corrections to the $t\bar{b}W$ coupling have the form
\beqa
\delta g^{W{\rm t}}_R  = 0, ~~& &~~ \delta g^{W{\rm g}}_R = 0, \nonumber \\
\delta g_L^{W{\rm t}} = -\frac{1}{4}\frac{\lambda_T^2 v^2 g^W}{m_T^2},
~~& &~~\delta g_L^{W {\rm g}} = \frac{v^2}{4 f^2}g^W s_\psi^2 \left(c_\psi^2-
s_\psi^2 -\frac{c_\psi^2 c_W^2}{c_W^2-s_W^2}\right),
\eeqa{delta_tbW}
where $g^W$ is the SM $t\bar{b}W$ coupling. These corrections induce a shift
in the top width, $\delta\Gamma_t/\Gamma_t=2\delta g_L^W/g^W$. The induced
shift, as a function of the angle $\psi$, is plotted in the right panel of
Fig.~\ref{fig:LH}, where the parameter $\lambda_T$ is varied between 0.5
and 2 for $m_T=0.5, 1.0$ and 2.0 TeV. The accuracy of the top width measurement
expected at the ILC~\cite{snowmass01} will allow to observe this effect in
most of the natural parameter space.

\if
The shift in the $t\bar{t}h$ coupling has the form
\beq
\frac{\delta\lambda_{t\bar{t}h}}{\lambda_{t\bar{t}h}}\,=\,
\frac{1}{3}\frac{v^2}{f^2}-
\frac{\lambda_T}{2}\frac{v^2}{fm_T},
\eeq{tth}
where the reference value is $\lambda_{t\bar{t}h} = \sqrt{2}m_t/v$,
and $v$ is inferred from the experimentally measured values for
$G_F$, $M_Z$ and $\alpha$. \draftnote{So far this assumes that $v$ is not
shifted, as would be the case in the T parity model; the extra correction
due to a shift in $v$ is to be computed.}
\fi

\section{LITTLEST HIGGS WITH T PARITY}

\begin{figure*}[t]
\centering
\includegraphics[width=60mm,angle=90]{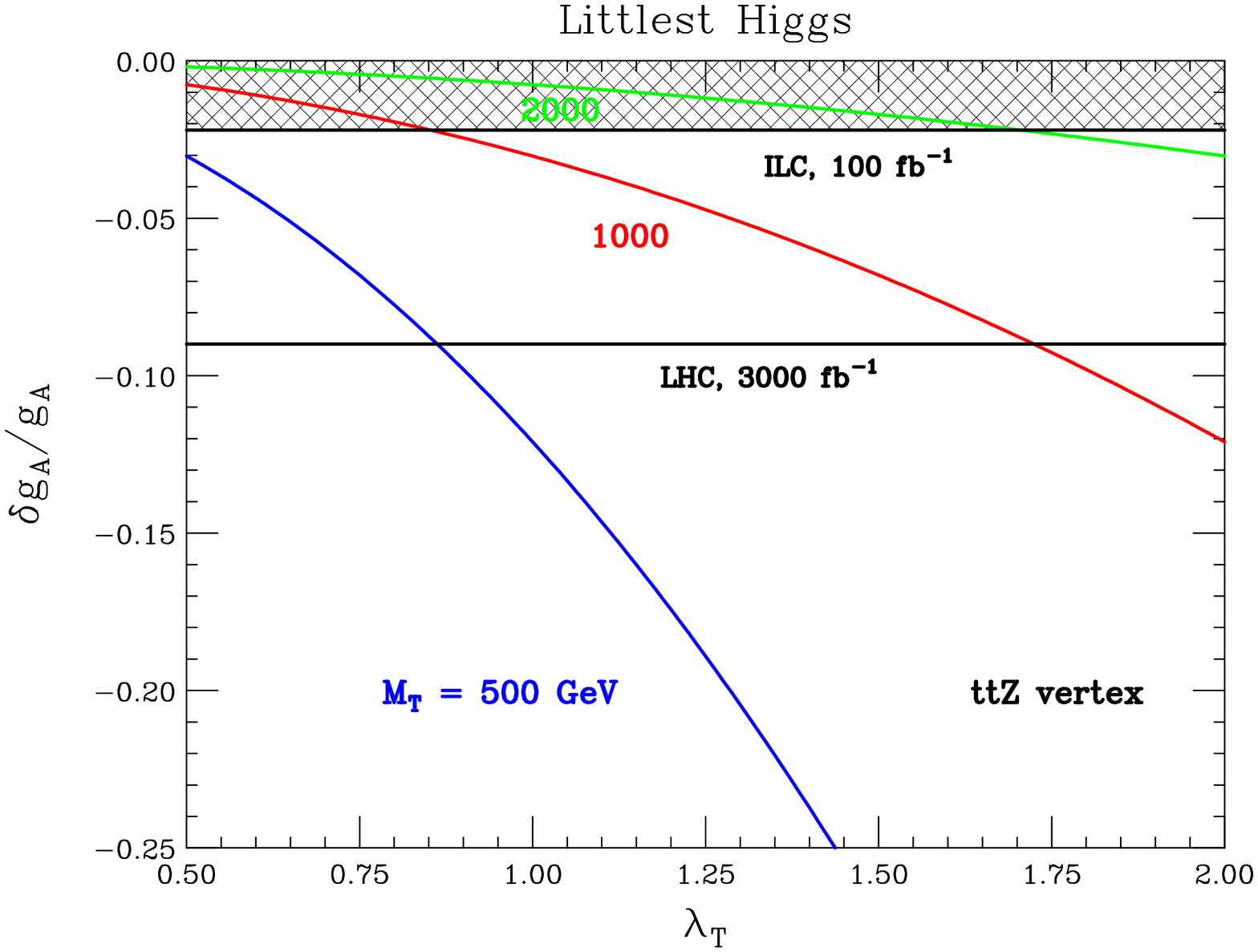}
\hskip1cm
\includegraphics[width=60mm,angle=90]{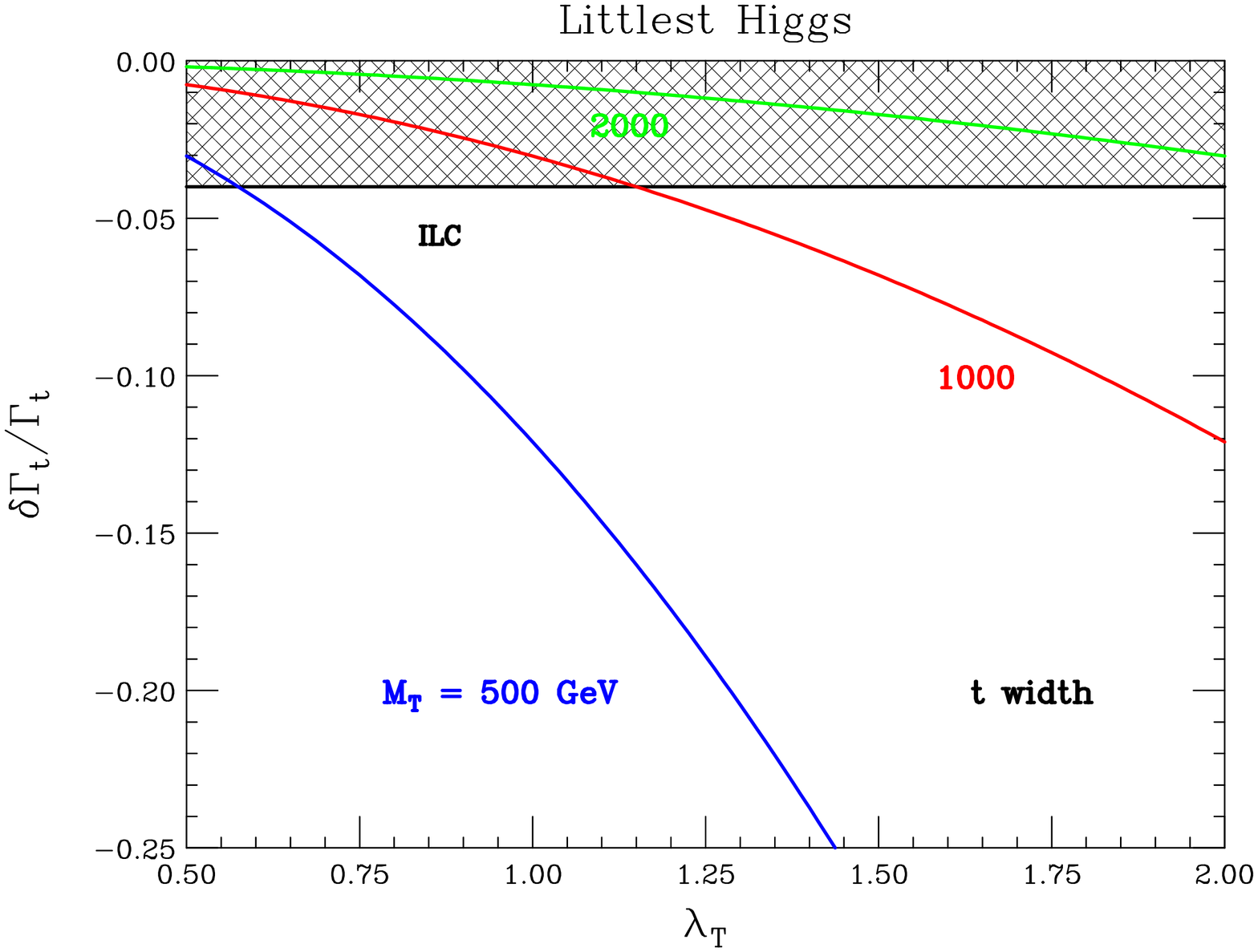}
\caption{The corrections to the $t\bar{t}Z$ coupling (left panel) and the top
quark width (right panel) in the $SU(5)/SO(5)$ Littlest Higgs model with
T parity. The regions in which the ILC would observe no 
deviation from the SM are shaded.}
\label{fig:Tpar}
\end{figure*}

The LH model can be extended to include a discrete symmetry, T
parity~\cite{LHT}, which greatly reduces the contributions to
precision electroweak observables~\cite{HMNP}. 
The main new feature in this model is the
absence of the gauge boson mixing, since light and heavy gauge
bosons have opposite charges under T parity. The top-heavy top
mixing is still present, however. The resulting corrections to the
$t\bar{t}Z$ and $t\bar{b}W$ vertices are identical to the
corresponding shifts in the model without T parity, $\delta
g^{Z{\rm t}}_L$ and  $\delta g^{W{\rm t}}_L$, given in
Eqs.~\leqn{delta_ttZ} and~\leqn{delta_tbW}. The shift in the axial
$t\bar{t}Z$ coupling is plotted in the left panel of
Fig.~\ref{fig:Tpar}. (The shift in the vector coupling is
identical up to a sign.) The correction to the top width is shown
in the right panel of Fig.~\ref{fig:Tpar}. Again, both effects
should be observable at the ILC.

\begin{acknowledgments}
The authors wish to thank Shrihari Gopalakrishna, JoAnne Hewett, 
Michael Peskin and Tom Rizzo for useful discussions.
CFB is supported by the US Department of Energy under
contract DE-AC02-76SF00515, MP is supported by the
NSF grant PHY-0355005, and FP is supported by
the University of Wisconsin Research Committee with funds granted by the
Wisconsin Alumni Research Foundation.
\end{acknowledgments}


\begin{thebibliography}{9}   

\bibitem{snowmass01}
  T.~Abe {\it et al.}  [American Linear Collider Working Group],
in {\it Proc. of the APS/DPF/DPB Summer Study on the Future of Particle Physics (Snowmass 2001) } ed. N.~Graf,
  arXiv:hep-ex/0106057.

\bibitem{review}
For a recent review and more references, see
M.~Schmaltz and D.~Tucker-Smith,
  arXiv:hep-ph/0502182.

\bibitem{PPP}
  M.~Perelstein, M.~E.~Peskin and A.~Pierce,
  Phys.\ Rev.\ D {\bf 69}, 075002 (2004)
  [arXiv:hep-ph/0310039].

\bibitem{ATLAS}
  G.~Azuelos {\it et al.},
  Eur.\ Phys.\ J.\ C {\bf 39S2}, 13 (2005)
  [arXiv:hep-ph/0402037].

\bibitem{littlest}
  N.~Arkani-Hamed, A.~G.~Cohen, E.~Katz and A.~E.~Nelson,
  JHEP {\bf 0207}, 034 (2002)
  [arXiv:hep-ph/0206021].

\bibitem{LHT}
  H.~C.~Cheng and I.~Low,
  JHEP {\bf 0309}, 051 (2003)
  [arXiv:hep-ph/0308199];
 I.~Low,
  JHEP {\bf 0410}, 067 (2004)
  [arXiv:hep-ph/0409025].

\bibitem{pew}
 C.~Csaki, J.~Hubisz, G.~D.~Kribs, P.~Meade and J.~Terning,
  Phys.\ Rev.\ D {\bf 67}, 115002 (2003)
  [arXiv:hep-ph/0211124];
  J.~L.~Hewett, F.~J.~Petriello and T.~G.~Rizzo,
  JHEP {\bf 0310}, 062 (2003)
  [arXiv:hep-ph/0211218].


\bibitem{LHC}
  U.~Baur, A.~Juste, L.~H.~Orr and D.~Rainwater,
  Phys.\ Rev.\ D {\bf 71}, 054013 (2005)
  [arXiv:hep-ph/0412021].

\bibitem{HMNP}
  J.~Hubisz, P.~Meade, A.~Noble and M.~Perelstein,
  arXiv:hep-ph/0506042.

\end{thebibliography}
\end{document}